\documentclass[cits]{PoS}
\usepackage{longtable}
\usepackage{lscape}
\usepackage{graphicx}

\title{Variability of low-luminosity AGNs: a simultaneous X--ray/UV look with {\it Swift}}

\ShortTitle{Low-Luminosity AGNs with {\it Swift}}

\author{ \speaker{P.\ Romano},$^a$ E.\ Pian,$^b$ D.\ Maoz,$^{cd}$ 
A.\ Cucchiara,$^e$ C.\ Pagani,$^{e}$ V.\ La Parola,$^a$  \\
\llap{$^a$}INAF, Istituto di Astrofisica Spaziale e Fisica Cosmica, \\
         Via U.\ La Malfa 153, I-90146 Palermo, Italy\\
\llap{$^b$}INAF--Osservatorio Astronomico di Trieste, Via G.\ Tiepolo 11, Trieste, I-34143, Italy \\
         via E.\ Bianchi 46, I-23807 Merate, Italy\\
\llap{$^c$}School of Physics and Astronomy, Tel-Aviv University, Tel-Aviv 69978, Israel\\
\llap{$^d$}INAF--Osservatorio Astrofisico di Arcetri, Largo E. Fermi 5, Firenze, I-50125, Italy\\
\llap{$^e$}Department of Astronomy \& Astrophysics, Pennsylvania State University, 
525 Davey Laboratory, University Park, PA 16802\\
E-mail: \email{romano@ifc.inaf.it}
}

\abstract{
We present the results of an investigation of the X-ray and UV
properties of four LINERs observed with {\it Swift}, aimed at constructing
good S/N and strictly simultaneous UV-X--ray SEDs. 
In the current paradigm, LINER emission is dominated by geometrically thick, radiatively
inefficient radiation flows (RIAFs) as opposed to radiatively efficient, 
geometrically thin accretion disks thought to power higher luminosity AGNs (Seyferts and QSOs).
However, some recent studies have found more similarities than
differences between the SEDs of LINERs and those of more luminous AGNs, suggesting
that LINERs are powered by the same mechanisms active in higher
luminosity AGNs. 
Our new observations allow us to test this idea. 
In particular, XRT affords long and sensitive monitoring of the X--ray emission.
We detect significant variability in M81 and, for the first time, in NGC~3998. 
The maximum amplitude variations over time scales of some hours are 30\,\% in both M81 and 
NGC~3998. NGC~3998 exhibits a variation of the same amplitude on a time scale of 9 days. 
M81 varies significantly over 2 years, with a maximum change of a factor 2 in 6 months.
The X--ray variability detected in 2 of our sources, and in particular in NGC~3998, 
puts into question the interpretation of their powering mechanism as an inefficient 
accretion flow, because one of the characteristics of this model is the lack of variability. 
The identification of NGC~3998 with a low power AGN appears more viable.
}

\FullConference{7th INTEGRAL Workshop\\
		 September 8-11 2008\\
		 Copenhagen, Denmark}

\begin{document}

\section{Introduction}

Low-luminosity AGNs are so common that virtually all massive galaxies 
may host some weak activity unrelated to stellar processes. 
Therefore, in order to understand super massive black hole activity in the local 
Universe and its evolution, it is necessary to study the demographics and 
physics of these objects. In particular, 
Low-Ionization Nuclear Emission Line Regions (LINERs) 
are low-luminosity active nuclei (AGNs) 
defined on the basis of their optical spectral line ratios
\cite{Heckman1980:linerdef,Ho2008:llagnannrev}, 
of which several subclasses may be defined according to their 
properties  \cite{Chiaberge2005:linersdual}.  The main open question 
is the origin of their optical spectrum and multi-wavelength emission 
in general: which fraction of the power source is non stellar and, 
if it is due to accretion, what is the accretion regime and the radiation conversion efficiency?     
It has been claimed that the lack of X-ray variability, the non detection of broad Fe K$\alpha$ lines
down to stringent limits \cite{Ptak2004:ngc3998}, and the weakness or absence 
of the characteristic ``big blue bump'' in their optical/near-ultra-violet (UV) spectra,
traditionally observed in Seyferts 
\cite{Quataert1999:linersriaf,Chiaberge2006:ngc4565}, indicate
that their engines may be intrinsically different from those of the more 
luminous AGNs and could consist of radiatively inefficient accretion flows.

However, X-ray observations of these sources have been so far sparse and not sensitive enough
to detect variability. Furthermore, ultra-violet (UV) monitoring with the 
{\it Hubble Space Telescope} (HST) Advanced Camera for Surveys (ACS) of 
a sample of 17 LINERs has revealed the presence of bright and variable UV nuclei \cite{Maoz2005}.  
\cite{Maoz2007:lowlumagns}, by coupling these UV measurements with non simultaneous 
X-ray measurements with {\it ASCA}, {\it Chandra}, and {\it XMM-Newton}, has demonstrated that the 
UV-to-X-ray flux ratios in LINERs are similar to those of Seyferts  (although the former sources 
are much less luminous than the latter).  
Therefore, contrary to the common paradigm, LINERs may be qualitatively not very 
different from Seyferts, but rather may be ``scaled-down'' analogues of them, with the
same emission mechanisms operating in both classes.  

Here we concentrate on LINER variability by obtaining, for the first
time, extended X-ray and UV datasets for a small sample of LINERs.    
{\it Swift} is well suited to this task because of its flexible and efficient
scheduling and because it can provide long, simultaneous and accurate UV and 
X-ray monitoring.

\section{Sample selection}

\setcounter{table}{0}
\begin{small}
\begin{table} 	
  \begin{center} 	
 \caption{XRT spectral fit parameters.  } 	
 \label{liners:tab:specfits} 	
 \begin{tabular}{lrrrrrrrr} 
 \hline 
 \hline 
 \noalign{\smallskip} 
 Name	& Distance & $N^{\rm G,a}_{\rm H}$	&$N^{\mathrm{b}}_{\rm H}$&		$\Gamma^{\mathrm{c}}$& 		$\chi^2_{\rm red}/$dof	&	$F^{\mathrm{d}}$ & $F^{\mathrm{d}}$\\
 	& (Mpc) & (cm$^{-2})$	& (cm$^{-2}$) 	&		   & &(0.2--1\,keV)	&  \\
  \noalign{\smallskip} 
 \hline 
 \noalign{\smallskip} 
M81      & 3.6 & 5.55	&$10.45_{-0.89}^{+0.93}$ &$2.04_{-0.04}^{+0.04}$ &$0.966/290$ &$11.2_{-0.1}^{+0.1}$ &$14.9_{-0.4}^{+0.4}$  \\ 
NGC~3998 & 13.1 & 1.01	&$ 6.85_{-1.20}^{+1.26}$ &$1.95_{-0.06}^{+0.06}$ &$0.948/188$ &$6.2_{-0.8}^{+0.1}$  &$9.8_{-0.4}^{+0.5}$   \\ 
NGC~4203 & 15.1 & 1.11	&$ 2.74_{-2.74}^{+4.51}$ &$1.81_{-0.21}^{+0.24}$ &$0.699/16$  &$1.4_{-0.4}^{+0.2}$  &$2.8_{-0.4}^{+0.6}$    \\
NGC~4579 & 21 & 2.97	&$ 7.01_{-1.34}^{+1.43}$ &$1.92_{-0.07}^{+0.07}$ &$0.972/145$ &$4.1_{-0.1}^{+0.1}$  &$6.8_{-0.4}^{+0.4}$   \\ 
 \noalign{\smallskip}
  \hline
  \end{tabular}
  \end{center}
  $^{\mathrm{a}}$  Hydrogen column densities derived from
    \cite{Kalberla2005:nhgalsur} in units of $10^{20}$ cm$^{-2}$ and consistent with the $A_B$ extinction 
values reported in \cite{Maoz2007:lowlumagns}, using a typical Milky
Way gas-to-dust ratio, $5 \times 10^{21}$ cm$^{-2}$ mag$^{-1}$. \\
   $^{\mathrm{b}}$ Hydrogen column densities from the XRT
      spectral fits in units of $10^{20}$ cm$^{-2}$. \\ 
        $^{\mathrm{c}}$  Photon index, $f_E \propto E^{-\Gamma}$. \\
       $^{\mathrm{d}}$  Unabsorbed flux in units of $10^{-12}$ erg cm$^{-2}$ s$^{-1}$.   \\         
  \end{table} 
 \end{small}

Our sample includes the brightest LINERs in the \cite{Maoz2007:lowlumagns} UV sample, 
M81, NGC~3998, NGC~4203, and NGC~4579, that are all Type I LINERs (i.e., with detected 
broad H$\alpha$ components) according to \cite{Ho1997:linerspec}.  
Data  for two of these sources, M81 and NCG~4203 were already available in the {\it Swift} archive:
M81 (net exposure of 33.7\,ks)  has an X-ray Telescope spectrum of
excellent quality and even the suggestion of an Fe K$\alpha$ line;   
NGC~4203  (5.3\,ks) has a good XRT spectrum, although slightly under exposed with 
respect to M81. 
NGC~3998 and NGC~4579 were observed with {\it Swift} for the first
time (27.4 and 20.8\,ks, respectively).  

\section{Data analysis}

\subsection{XRT}

The XRT data were processed with standard procedures ({\sc xrtpipeline} 
v0.11.4), filtering and screening criteria by using {\sc  FTOOLS} in the
{\sc Heasoft} package (v.6.3.1). 
We extracted spectra  for each XRT observation, as well as for the cumulative 
observing campaigns. 
We also extracted light curves in the 0.2--10\,keV (total), 
0.2--1\,keV (soft, S) and 1--10\,keV (hard, H)  bands. 
The host galaxy contribution in the XRT extraction region was evaluated 
using archival {\it Chandra} images of these LINERs obtained with
ACIS-S with the longest exposure and was found to  range 
from 6\,\% (NGC~3998) to 16\,\% (NGC~4203), and was subsequently
ignored.
We fit the mean XRT spectrum for each object with a simple absorbed power-law
model with free absorption and photon index (see Table~\ref{liners:tab:specfits}). 
The spectra show no significant
absorption features superimposed on the power-law continuum. 
Although the uncertainties on the fitted Hydrogen column densities
(Table~\ref{liners:tab:specfits}, Col.\ 4)
are not small, all sources show significant evidence for an absorption in excess of
the Galactic one (also reported in Table~\ref{liners:tab:specfits}).

\subsection{UVOT}

The {\it Swift} UltraViolet-Optical Telescope 
observed the four targets with the filters $u$ (3465\,\AA)
$uvw1$ (2600\,\AA), $uvm2$ (2246\,\AA), $uvw2$ (1928\,\AA).  
The filter choice was driven by the objective of maximizing 
the nuclear signal, which is dominant in the UV,  while 
minimizing the emission from the bulges of the galaxies.
After verifying that the UVOT counts show no significant variability,
we have coadded the images for each object.
The data analysis was performed using the {\sc uvotsource} 
task included in the latest {\sc Heasoft} software. 

Using archival HST images of 
these LINERs at 2500\,\AA\ and 3300\,\AA\  obtained with the High
Resolution Channel (HRC) of the HST ACS \cite{Maoz2005}, we evaluated the host 
galaxy contribution for each object in a 2\,arcsec-radius circular area centered on 
the nucleus.
The resulting magnitudes were converted into fluxes using the 
latest in-flight flux calibration factors and zero-points \cite{Poole2008:uvotconverfac}.
The host galaxy contribution is dominant in the $u$ band (generally comparable to 
the total flux observed by UVOT; sometimes it exceeds it by 20 to 50\,\%) and it  decreases
almost steadily toward  the shorter wavelengths.  It is minimal in the $uvw2$ filter 
(1928\,\AA) where it contributes at most 60\,\% of the observed flux (NGC~3998). 



\subsection{Results}

\begin{figure*}
	\vspace{-2.0cm}
        \hspace{-1.5truecm}
 	\includegraphics[angle=0,width=17cm,height=21cm]{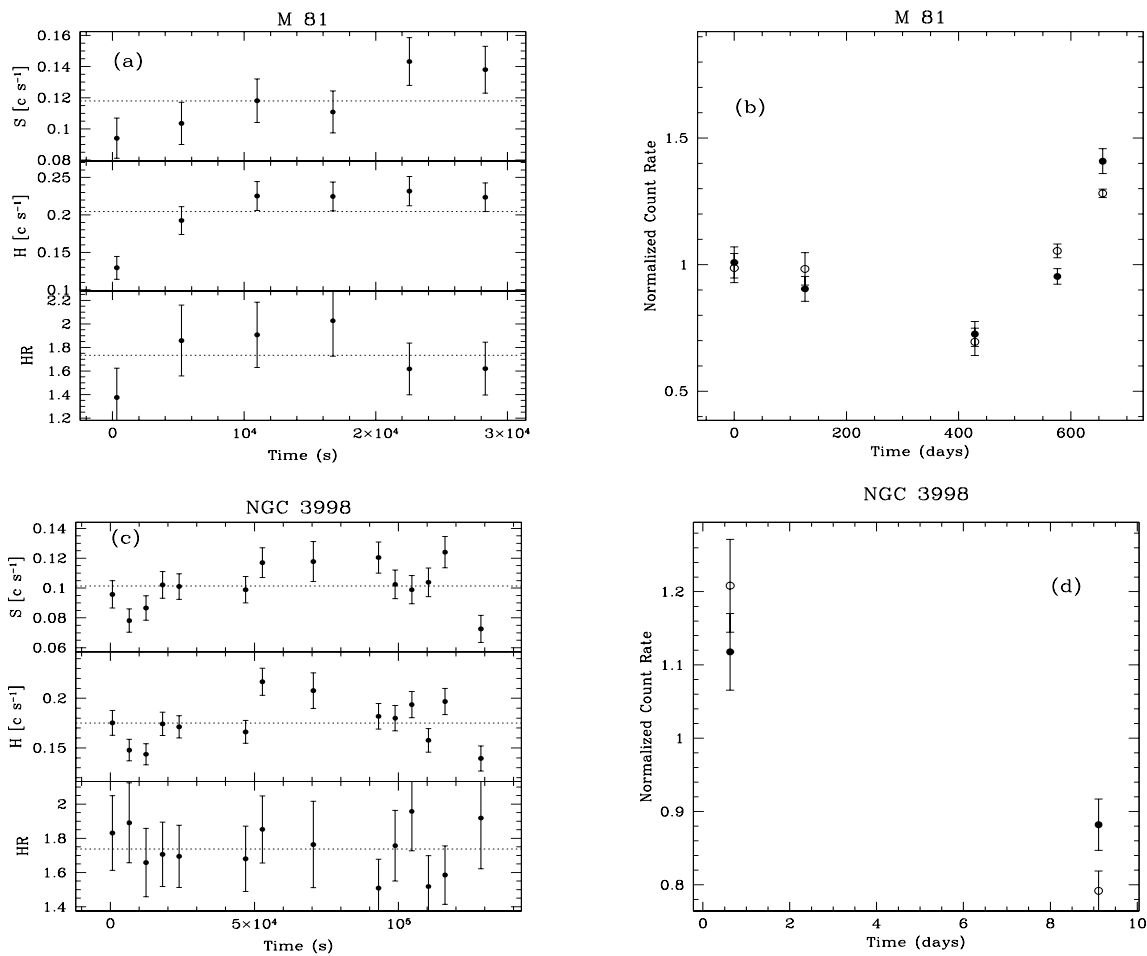}
	\vspace{-4.0cm}
	\caption{   {\it Swift}/XRT background-subtracted light curves
          (count rate in counts s$^{-1}$) and hardness ratios (i.e., 1--10 keV to 0.2--1 keV flux ratios):
    {\it (a)}  0.2--1\,keV (top), 1--10\,keV (middle) light curves and
   hardness ratio curves (bottom) of M81 starting on 2006 June 24.004
   UT. The binning time interval corresponds to the orbit duration. 
{\it (b)} 0.2--1\,keV (filled circles) and 1--10 keV (open circles)
light curves of M81 between 2005 and 2007. 
Each curve is normalized to its average (0.16 and
   0.28 counts s$^{-1}$ for the 0.2--1\,keV and 1--10\,keV curves, respectively). Each point is the average
   of the flux measured during each of the 5 pointings in that given
   band. The time origin corresponds to 2005 Apr 21.0 UT.    
{\it (c)} 0.2--1\,keV (top), 1--10\,keV (middle) light curves and
hardness ratio curves (bottom) 
of NGC~3998 starting on 2007 April 29.11 UT. The binning time interval corresponds to the orbit duration.
{\it (d)} 0.2--1\,keV (filled circles) and 1--10 keV (open
        circles) light curves of NGC~3998 
averaged over each of the two pointings in April 2007. Each curve is normalized to its average (0.11 and
   0.21 counts s$^{-1}$ for the 0.2--1\,keV and 1--10\,keV curves, respectively).
    The time origin  corresponds to 2007 Apr 20.0 UT.  
	}
\label{liners:fig:xlc_all}
\end{figure*}

Significant X--ray variability is detected in NGC~3998 for the first time. 
The variations are about 30\,\% on both intraday and interday time scales. 
In Fig.~\ref{liners:fig:xlc_all}c is reported the light curve of one of the 
2 {\it Swift} pointings we have made and in Fig.~\ref{liners:fig:xlc_all}d are reported 
the average fluxes of the 2 pointings, in soft and hard X--rays. 
This is unprecedented for this source, that has always been reported not to be 
variable in the X--rays \cite{Pellegrini2000b:ngc3998bsax,Ptak2004:ngc3998}. 
M81 varies with a maximum amplitude of 30\,\% on intra-day time-scales. 
In Fig.~\ref{liners:fig:xlc_all}a we report the light curve of one of our 
5 observations, in which both the soft and hard X--rays are seen to vary 
in a correlated way, although the significance of the soft X--ray variation is limited. 
The month time scale variability has a larger amplitude, up to a factor 2 in 6 months 
in both soft and hard X--rays, and good correlation between the 2 bands (Fig.~\ref{liners:fig:xlc_all}b). 
This is similar to previous findings; indeed, earlier X--ray observations 
of M81 had reported variations of even a factor 4 \cite{LaParola2004:CXOm81}.
Variability on a minute time scale is seen of up to 30--40\,\%, but it occurs
on low states of the sources, and therefore we consider it highly dubious.

NGC~4579 and NGC~4203  (Fig.~\ref{liners:fig:xlc_all}e, f), on the other hand, 
exhibit no significant variability on any time scale. 
One reason may be that our monitoring was carried out for too short a baseline, 
over which there was no chance to detect variations, that occur stochastically. 

After host galaxy subtraction, we do not detect significant optical/UV 
variability in any of our sources.

\subsection{Conclusions}

The variability we find is not dissimilar from that of Seyferts.
While this does not prove that LINERs have (weak) AGNs at their centers as opposed 
to advection dominated accretion flows (ADAF), it puts the issue back into question. 
In particular, we detect significant variability in an object (NGC~3998) that had been 
considered a very good example of an ADAF, based on the lack of variability 
\cite{Pellegrini2000b:ngc3998bsax,Ptak2004:ngc3998}. 

In summary, we suggest that, if monitored with sufficient accuracy, long time and on a wide 
band, some LINERs can vary, albeit with modest amplitude. 
This is not inconsistent with the variations of Seyferts. 
A real test of the emission model calls for a multiwavelength approach. 
Our UV observations are of limited help, because they are affected by the host galaxy 
and are therefore not sufficiently accurate to study UV variability. 
A compelling test would consist in strictly simultaneous XRT and radio observations, 
possibly accompanied by sensitive HST observations in the UV.
In a forthcoming paper we will present the results of the analysis of the simultaneous 
X-ray and UV spectral energy distributions \cite{Romano2009}.

\vfill

\acknowledgments
We thank N.\ Gehrels for approving this set of ToOs and the 
{\it Swift} team, in particular the duty scientists and science planners, 
for making these observations possible. 
P.R.\ thanks INAF-IASF Milano, where part of the work was carried out, 
for their kind hospitality.
This research has made use of NASA's Astrophysics Data System Bibliographic Services,  
as well as the NASA/IPAC Extragalactic Database (NED), which is operated 
by the Jet Propulsion Laboratory, California Institute of Technology, under contract with 
the National Aeronautics and Space Administration.


\setcounter{figure}{0}

\begin{figure*}
	\vspace{-8.0cm}
        \hspace{-1.5truecm}
 	\includegraphics[angle=0,width=17cm,height=21cm]{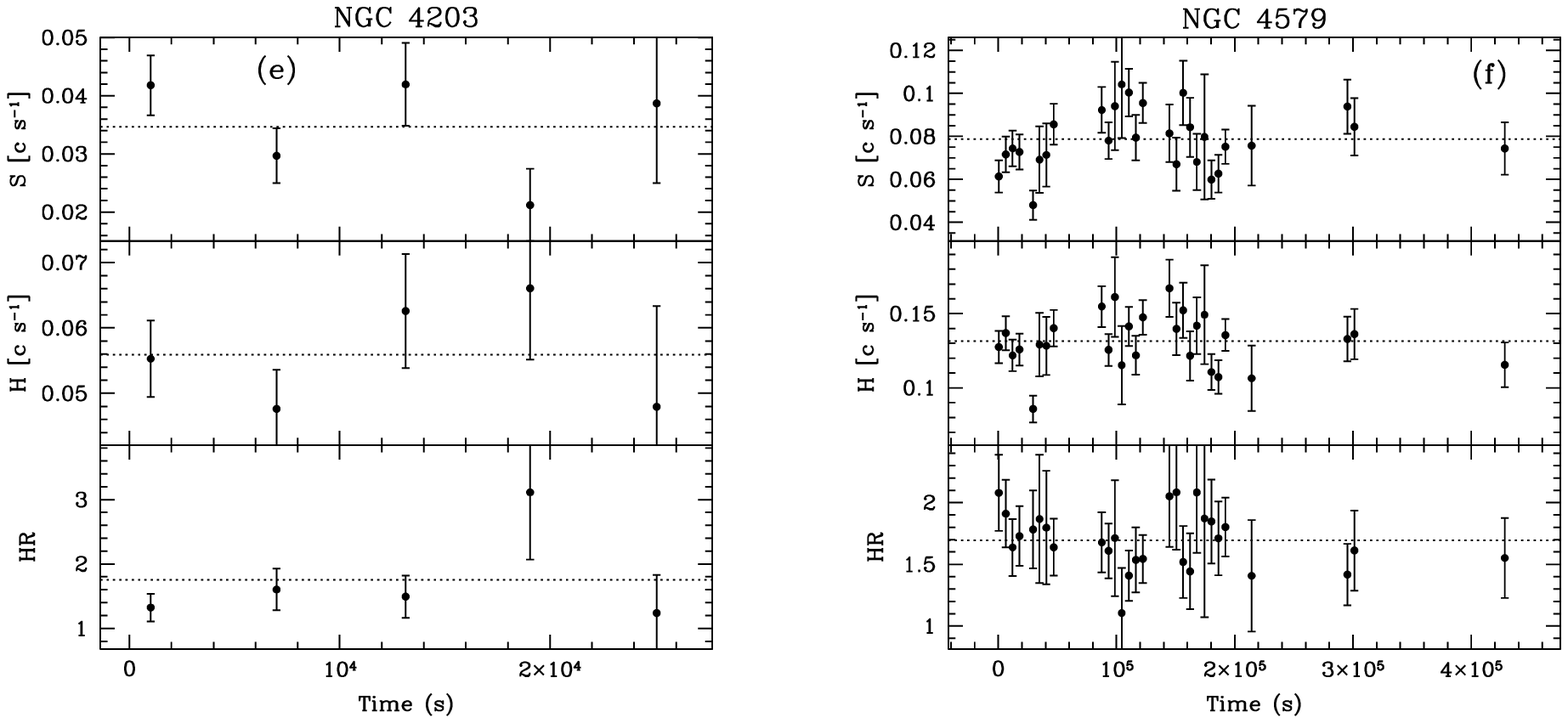}
	\vspace{-8.0cm}
	\caption{ 
{\it (e)} 0.2--1 keV (top), 1--10 keV (middle) light curves and hardness ratio 
curves (bottom) of NGC~4203 starting on 2005 December 25.004 UT. 
The binning time interval is the orbit duration. 
{\it (f)} 0.2--1 keV (top), 1--10 keV (middle) light curves and hardness 
ratio curves (bottom) of NGC~4579 starting on 2005 May 15.04 UT. 
The binning time interval is the orbit duration. 
	}
\end{figure*}

\end{document}